\documentclass[ngerman, italian, english, UKenglish]{scrreprt}
\usepackage[latin9]{inputenc}
\usepackage{geometry}
\usepackage{fancyhdr}
\usepackage{float}
\usepackage{units}
\usepackage{amsmath}
\usepackage{amssymb}
\usepackage{graphicx}
\usepackage{subfigure} 
\usepackage{caption}
\usepackage{textcomp}
\usepackage{setspace}
\usepackage{tikz}

\usepackage{hyperref}

\makeatletter

\usepackage{enumitem}

\usepackage{tikz}
\usetikzlibrary{scopes}
\usepackage{marvosym}
\usetikzlibrary{shapes,snakes}
\usepackage{fancyhdr}
\usepackage{chngcntr}
\usepackage{wallpaper}
\usepackage{multicol}
\usepackage{upgreek}
\usepackage{url}
\usepackage{pdflscape}
\usepackage[exponent-product = \cdot, output-decimal-marker={,},load-configurations=abbreviations]{siunitx}
\makeatother

\usepackage{babel}

\newcommand{\Dd}{\mathrm{d}}
\newcommand{\be}{\begin{equation}}
\newcommand{\ee}{\end{equation}}
\newcommand{\bea}{\begin{eqnarray}}
\newcommand{\eea}{\end{eqnarray}}

\counterwithout{equation}{chapter}
\counterwithout{table}{chapter}
\counterwithout{figure}{chapter}

\begin{document}
\setcounter{figure}{0}
\setcounter{equation}{0}
\setcounter{table}{0}
\setcounter{page}{37}
\newpage
\section*{Gravitational waves and cosmic expansion: similarities and differences}
\addcontentsline{toc}{section}{Gravitational waves and cosmic expansion (M. Pössel, Heidelberg)}

Markus P\"ossel, Haus der Astronomie and Max Planck Institute for Astronomy, Heidelberg

\bigskip{}

\noindent \emph{Gravitational waves and cosmic expansion are both described in terms of Einstein's general relativity. This article explores the similarities between the two phenomena, as well as some differences, using the fundamental concept of the metric of spacetime, expressed via a line element. In both contexts, the focus is on the metric description of space, as opposed to the components of the metric determining the properties of time. This allows for a simplified, more accessible discussion.}

\setlength{\columnsep}{0.75cm}
\begin{multicols}{2}
\noindent The main focus of this summer school is on gravitational waves.\footnote{This text has been published in K.-H. Lotze \& Stefan V\"olker (eds.), {\em Proceedings of the Heraeus Summer School ``Astronomy from 4 Perspectives: Gravitational Wave Astronomy''} (Jena, 31 Aug--5 Sep 2015), pp. 37--48.} But in order to understand how these waves interact with matter (including gravitational wave detectors), it is helpful to take a broader view that includes not only gravitational waves but another phenomenon intimately connected with the relativistic description of our universe: cosmic expansion as the basis for our current models of cosmology. Both phenomena are described in terms of the {\em spacetime metric} encoding the geometry of space and time. We begin by introducing the concept of spacetime metrics over the next three sections.

\subsubsection*{The three--fold role of coordinates}

\noindent Usual Cartesian coordinates in physics serve a threefold purpose. For one, they provide a scheme for naming points in space. Once you have introduced your Cartesian coordinate system, it is as if every point in space were sporting a little name tag, bearing a unique name such as $(1.0,-2.6,9.1)$ or similar. The unique name can be used to refer to that specific point, and that point only.

In addition, the coordinate values encode some information about proximity. In the Cartesian plane, the point $(0,0.5)$ is closer to the point $(0,0.4)$ than to the point $(0,9.5)$. Finally, Cartesian coordinates allow for a direct computation of distances between points. Given two points and their coordinates, say $P_1=(x_1,y_1,z_1)$ and $P_2=(x_2,y_2,z_2)$, their distance $s$ is given by 
\be
s = \sqrt{ (x_1-x_2)^2 + (y_1-y_2)^2 + (z_1-z_2)^2},
\ee
corresponding to a three-dimensional version of Pythagoras' theorem. By introducing a shorthand notation  $\Delta x = x_1-x_2$ and analogous expressions for the other coordinates, the same relation can be written more compactly as
\be
s^2 = \Delta x^2 + \Delta y^2 + \Delta z^2.
\ee
An additional bonus is that when we plot points in Cartesian coordinates, say the points in a plane, we usually make sure to draw all distances faithfully. Hence, in the usual xy coordinate system, drawn on a piece of paper, we can simply use a ruler to directly measure the distances between points, and rest assured that calculations using point coordinates will yield the same result (bar scale factors linking the scale of our drawing and the scale of the ruler). This is, of course, a teaching tool that is used extensively in a school setting. 

\begin{figure}[H]
\begin{center}
\includegraphics[width=1\columnwidth]{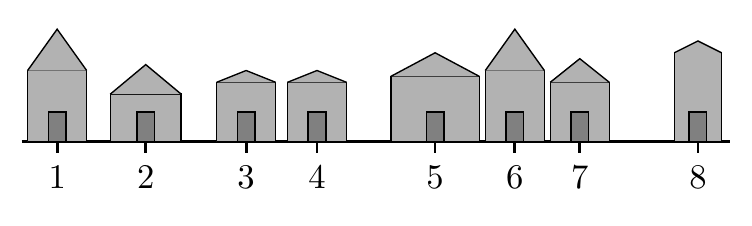}
\caption{Houses, numbered with integers}
\label{fig:HouseRow}
\end{center}
\end{figure}

For more general spaces and surfaces, and even for non-Cartesian coordinates in regular, Euclidean space or on a Euclidean space (such as spherical coordinates or polar coordinates), the relationship between distances and coordinate values is more complex. A simple, one-dimensional example is that of a row of houses, numbered with integers, as sketched in figure \ref{fig:HouseRow}.

To begin with, the house numbers are merely markers for each house; in our simple sketch, they have been associated with the center of the house, determined along the one-dimensional street. We can extend these numbers to a one-dimensional coordinate system by subdividing evenly the distance between one marker and the next; the point half-way between house no.\ 4 and house no.\ 5, for instance, would be the point with coordinate $4.5$, and the point three-quarters of the way between houses 2 and 3 would have $2.75$ as its coordinate.

Coordinates defined in this way perform two of the three roles we identified for Cartesian coordinates: they uniquely identify each point along the street, and they encode proximity: regarding house no. 3, house no. 5 is further away than no. 4. 

What the coordinates do not allow without additional information is f a direct calculation of distances. To calculated distances, we need information about the distances between the houses -- how far away is the mark in front of no. 1 from that in front of no. 2? How far away is the no. 2 mark from the no. 3 mark, and so on? If we know that, say, the no. 4 and no. 5 marks are $6.8\;\mbox{m}$ apart, then we know that our point $4.5$ is $3.4\;\mbox{m}$ from the mark no. 4, and the same distance from the no. 5 mark.

With this additional information, we can use our coordinates to compute distances between points along the street -- as long as we have that additional input, the information about how far the houses are apart. This additional information is our first example of a {\em metric}: a set of information that allows you to translate coordinate differences into distances. Let us look at a more general, two-dimensional example.

\subsubsection*{Line element of a 2-dimensional surface}
\label{CurvedSurface}

\noindent Consider an idealized rocky landscape, with hills and valleys, perfectly smooth and polished, without any breaks or sharp edges. That landscape is the stand-in for a general two-dimensional surface. Imagine that we draw a two-dimensional Cartesian coordinate system onto a sufficiently large rubber sheet. We do not just draw the axes, but all coordinate lines: the lines corresponding to $x=1$, $x=1.1$, $x=1.2$ and so on, and corresponding lines for $y=0.5$, $y=0.51$, and $y=0.52$ and many, many more. In reality, we can only draw a finite grid of lines, of course; in our thought experiment, we could imagine we had drawn {\em all} the lines. We also label all the lines, so whenever we see a line on that rubber sheet, we can identify which line $x=const.$ or $y=const.$ we are looking at. 

Each point on the rubber sheet has a unique coordinate label. After all, that point will be situated on exactly one line $x=const.$ and on one line $y=const.$; those two lines define that point's pair of coordinate values $(x,y)$. 

Now imagine that we spread the rubber sheet across our rocky landscape, making sure the sheet covers the surface tightly, with no pockets of air in between, and no wrinkles, each part of the sheet covering a corresponding part of the landscape. Evidently, there will be many places where we will need to stretch the rubber sheet to make sure it fits the surface snugly. Our coordinate lines will become general, curved lines on the surface. 

Even the distorted rubber sheet is sufficient to define a coordinate system on our hilly surface. After all, distortion and stretching do not change the fact that every point of the surface will have one x coordinate line and one y coordinate line intersecting at exactly that point on the rubber sheet. Every point of the surface has an $(x,y)$ coordinate pair.

\begin{figure}[H]
\begin{center}
\includegraphics[width=1\columnwidth]{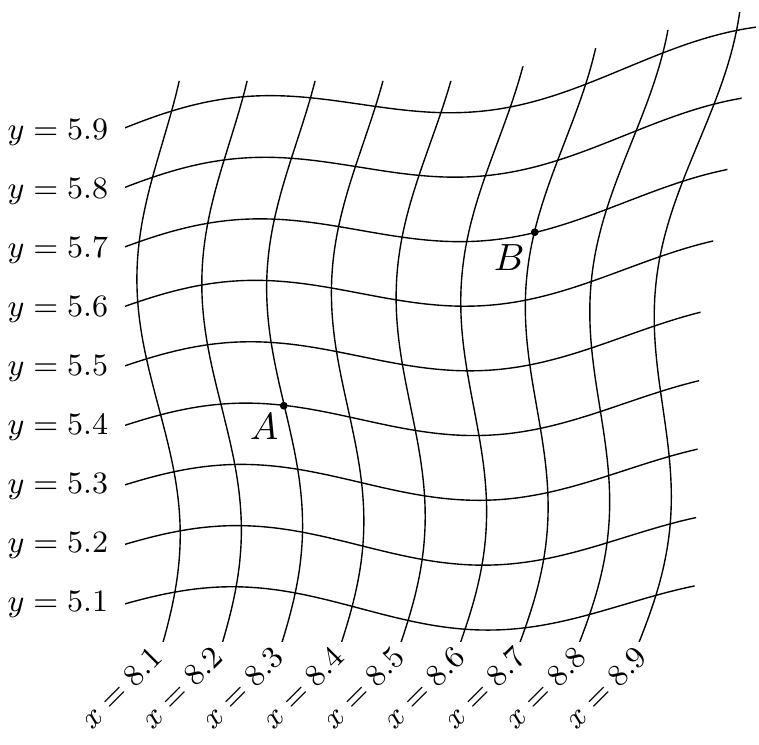}
\caption{The rubber sheet coordinate system on our surface, as seen by a birds-eye observer, showing selected coordinate lines $x=const.$ and $y=const.$}
\label{WavyLines}
\end{center}
\end{figure}

When we take a step back to look at our rubber-sheet-covered landscape, the coordinate lines will in general look wavy and distorted. Figure \ref{WavyLines} shows an example of what we might see, at least for a selected few coordinate lines. For the lines shown, you could immediately find the point $A=(8.3,5.4)$ or the point $B=(8.7,5.7)$, or any of the nearly 80 additional points lying on the intersection of the visible coordinate lines in figure \ref{WavyLines}. If a finer grid were shown, you could find many more points; an idealized, infinitesimally fine grid, including all coordinate lines, would allow you to find coordinates for every point in the region visible in figure \ref{WavyLines} by looking at the coordinate lines intersecting at that particular point.

One glance at figure \ref{WavyLines} will show you that this is not a Cartesian coordinate system on a plane surface. The wavy lines are a dead give-away. (Note, though, that simply by looking at the lines in this way, you could not tell whether this was a distorted rubber sheet on a plane, or on a more complex surface!)

But what happens when you zoom in on one particular point of the surface and the portion of rubber sheet covering the region around it? There is a more familiar case of such a zooming-in, namely the infinitesimal straightness that emerges when you look at a particular point on a differentiable curve plotted in the usual xy coordinate system: zoom in sufficiently, and the portion of the curve you are looking at will increasingly resemble a straight line. This is the visual motivation for defining the derivative of that function, which is closely related to the line you'd get if you had an infinitely strong zoom, letting you see the infinitesimal neighbourhood of your chosen point.

\begin{figure}[H]
\begin{center}
\includegraphics[width=1\columnwidth]{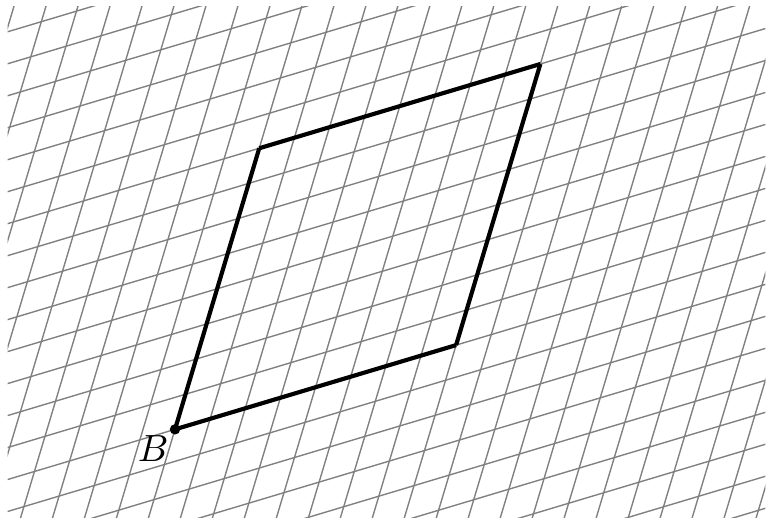}
\end{center}
\caption{A small neighbourhood around the point $B=(8.7,5.7)$ on the smooth surface in figure \ref{WavyLines}, sufficiently magnified for the coordinate lines to look like segments of straight lines.}
\label{InfinitesimallyStraight}
\end{figure}

The properties of our rocky surface (smooth, no breaks or sharp edges) and the procedure that gave us the coordinate lines (distorting the rubber sheet to fit the surface snugly) suggests that these lines, too, are differentiable. By that criterion, when we zoom in sufficiently close, we should be able to make the criss-crossing, general/curved coordinate lines look more and more like straight line segments. This suggests that, at suitably high magnification, the region around each point should look as shown in figure \ref{InfinitesimallyStraight}.

In particular, the straight line segments mesh together just as one would expect in a plane with Euclidean geometry. Our two-dimensional version of infinitesimal straightness thus reads: Zoom sufficiently close to a smooth and possibly curved surface, and the small (in the limit: infinitesimal) region you are looking at will be indistinguishable from a small subset of a Euclidean plane. 

We are using this property of curved surfaces implicitly whenever we consult a common road-map -- which, after all, maps a small portion of the surface of a sphere, namely the globe, onto a flat piece of paper. 

Let us focus on the area delineated by the parallelogram marked in figure \ref{InfinitesimallyStraight}. If we approximate the small region in question as a small subset of a Euclidean plane, then the coordinate lines on that plane define a coordinate system that is almost, but not quite Cartesian: the coordinate lines are indeed straight lines, but the x and y axis are not orthogonal to each other. 

\begin{figure}[H]
\begin{center}
\includegraphics{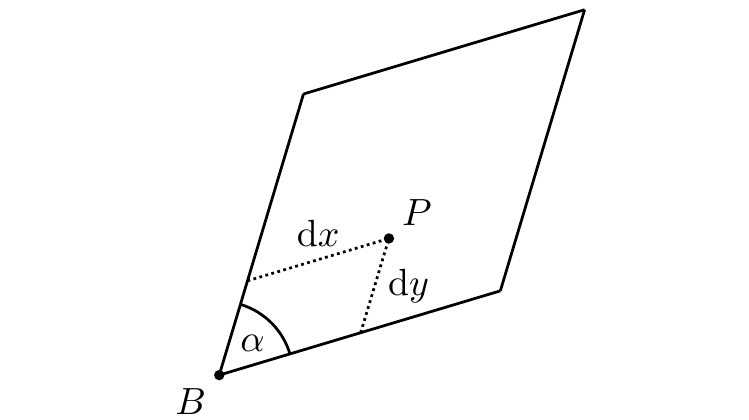}
\caption{Parallelogram adjoining the basepoint $B$. What is the distance between $B$ and some point $P$ whose coordinates are shifted against those of $B$ by $\Dd x$ in the x direction, $\Dd y$ in the y direction?}
\label{ParallelogramGeo}
\end{center}
\end{figure}

This need not keep us from calculating the distance of any point $P$ in the parallelogram from the basepoint $B=(8.7,5.7)$, though. The geometry of the situation can be seen in figure \ref{ParallelogramGeo}. More generally, let us place our basepoint at $B=(x,y)$. What do we need to know in order to determine distances? First of all, just as in the example with the house numbers, we need a scale factor. Consider a small shift $\Dd x$ in the x coordinate, moving us from $P_1=(x,y)$ to $P_2=(x+\Dd x,y)$. Define the scale factor $a$ linking the coordinate distance $\Dd x$ and the distance $\overline{P_1P_2}$ by 
\be
\overline{P_1P_2} = a\,\Dd x.
\ee
Analogously, we define a scale factor $b$ for small shifts in the y direction. At $B$, the angle between the x and y coordinate line is $\alpha$. From the law of cosines, it follows that for a general coordinate shift $(\Dd x,\Dd y)$, the distance $\Dd s$ between $B=(x,y)$ and the point $P=(x+\Dd x,y+\Dd y)$ is given by 
\bea
\nonumber
\Dd s^2 &=& a^2\,\Dd x^2 + 2 \,ab\cos(\alpha)\,\Dd x\,\Dd y + b^2\,\Dd y^2  \\[0.5em] 
&\equiv& g_{11}\,\Dd x^2 + 2 g_{12}\,\Dd x\,\Dd y + g_{22}\,\Dd y^2,
\eea
where the expression in the second row serves to define the {\em metric coefficients} $g_{11}, g_{12}$, and $g_{22}$. Such an expression linking infinitesimal distances $\Dd s$ and coordinate shifts $\Dd x, \Dd y, \ldots$ is called a {\em line element}. 

Just as in Pythagoras' theorem, this expression is second order in the coordinate shifts. The information linking coordinate shifts and distances is encoded in the metric coefficients. Taken together, the coefficients form the {\em metric} at that particular point. At other locations on the surface, there is an analogous formula linking coordinate shifts and infinitesimal distances; in general, the metric coefficients  $g_{11}, g_{12}$, and $g_{22}$ will take on other values than at our original basepoint $B$ as we move to different locations. In other words: the metric coefficients are functions of position, $g_{11}(x,y), g_{12}(x,y)$, and $g_{22}(x,y)$, and so is the metric. 

Once we know these functions, we can answer all questions about the {\em inner geometry} on the surface. In short, the inner geometry deals with all questions that a hypothetical two-dimensional being that is living on the surface can ask about curves in the surface, the geometric objects that can be constructed from such curves, and the various distances and angles involved. The answers, arrived at using the metric, could each be checked by appropriate measurements taken on the surface. 

On the other hand, the metric cannot tell us anything about the {\em exterior geometry}, that is, the specifics of how the two-dimensional surface is embedded in three-dimensional space. For a simple example, imagine all the ways one can embed a two-dimensional sheet of paper in space; the geometry of triangles etc. on the sheet remains Euclidean even when the embedding changes. In fact, formulating geometry in terms of a metric provides the means to discuss the geometry of curved surfaces without the need for such embeddings! 

The metric coefficients introduced here are the central elements of the general description for the inner geometry of smooth surfaces, which was found by Carl Friedrich Gau\ss\ (1777--1855). Bernhard Riemann (1826--1866) generalized this description to spaces of arbitrary dimension.

\subsubsection{Spacetime metric}

\noindent The properties of special relativity can be written using a metric, as well. At first glance, the line element looks a bit weird. While it is written as a square, the line element can become negative! In the usual coordinates of an inertial system $S$, using Cartesian coordinates for space, it can be written as 
\be
\Dd s^2 = -c^2\,\Dd t^2 + \Dd x^2 + \Dd y^2 + \Dd z^2, \label{SRTMetrik}
\ee
where $c$ is the speed of light in vacuum. All the usual special-relativistic effects, such as time dilation and length contraction, can be derived from this metric, which is known as the {\em Minkowski metric}.

The starting point for general relativity is a generalization of the metric (\ref{SRTMetrik}). The most general form of the metric will contain all second-order products of coordinate differences (such as $\Dd x\cdot\Dd t$ or $\Dd x\cdot\Dd z$, among others), and the metric coefficients can be functions of all the four coordinates.
 
For our simple analysis of cosmological and gravitational wave spacetimes, two general properties of such general spacetime metrics are sufficient: If we confine ourselves to $\Dd t=0$, then the remaining non-zero part of the metric can be used in direct analogy to what we did with the metric of our two dimensional surface to relate spatial coordinate shifts to spatial distances.\footnote{There is a caveat when it comes to interpreting these spatial distances; in general, the interpretation depends on the meaning of the time coordinate that was chosen -- in line with the fact that already plays an important role in special relativity, namely that measuring distances depends on one's notion of simultaneity.} 
 
The second general property is that the trajectories $x(t), y(t), z(t)$ taken by light always satisfy $\Dd s^2=0$, more specifically: if $\Dd x, \Dd y, \Dd z$ are the shifts in the spatial coordinate of a photon as the coordinate time interval $\Dd t$ passes, the line element $\Dd s^2$ for this particular set of coordinate shifts $\Dd x, \Dd y, \Dd z, \Dd t$ must vanish. It is straightforward to see that, in the case of (\ref{SRTMetrik}) in special relativity, this corresponds to light moving with the usual speed of light $c$.\footnote{
For the easiest way to see this, restrict yourself to light propagation in the x direction only. Then $\Dd s^2=0$ translates to $|\Dd x/\Dd t| = c$. 
}

In general relativity, the geometry of spacetime (encoded by the functions that specify the metric coefficients) is determined by the {\em Einstein Field Equations} (EFE). The EFE are second order differential equations linking the first and second derivatives of the metric coefficients with the mass, energy, momentum and pressure of whatever matter is present in the spacetime in question, which serves as a source of gravity. \footnote{An elementary treatment can be found in \cite{Baez2005}} A set of matching metric coefficients and suitable source terms, linked by the EFEs, is called a {\em solution} of the EFEs. All models of physical situations in the framework of general relativity are formulated in terms of suitable solutions. 

Using the EFEs, it is possible to calculate in a systematic manner the deviations from classical Newtonian gravity which occur in situations with comparatively weak gravity. Newtonian gravity itself can be described, in a coordinate system close to that chosen in classical physics, as a location-dependent coefficient $g_{00}$ in front of the metric's $\Dd t^2$ term.

\subsubsection{Metrics for expanding universes and for gravitational waves}
The simplest metric for a homogeneous and isotropic expanding universe, and the metric on which current cosmological models are based, is
\be
\Dd s^2 = -c^2\,\Dd t^2 + a(t)^2\left[\Dd x^2 +\Dd y^2 +\Dd z^2 \right]. \label{FLRW-Metrik}
\ee
In general relativity, as more generally in scientific modelling, choosing suitable coordinates is very important. In the case of (\ref{FLRW-Metrik}), the coordinates have been chosen adapted to the situation as follows: consider idealized galaxies that follow cosmic expansion without any additional velocity components (``peculiar velocity'' due to motion within galaxy clusters). Such galaxies are said to move with the {\em Hubble flow}; in our coordinate system, they are assigned constant coordinate values $x,y,z$. (This is known as using {\em comoving coordinates} -- the coordinate system moves along with the galaxies.) The time coordinate $t$ is measured by clocks moving alongside galaxies in the Hubble flow. These clocks are synchronized in exactly the right way that, for any fixed time $t=const.$, the local average density of the universe is the same at each location. (In other words: our notion of simultaneity is adapted to the homogeneity of the universe.)

This is the simplest case of {\em Friedmann-Lema\^{\i}tre-Robertson-Walker} universe (FLRW universe), namely an expanding universe with Euclidean spatial geometry (in the lingo of FLRW solutions, $k=0$; the other possibilities for FLRW universes are hyperbolic geometry $k=-1$ and spherical geometry $k=+1$).

The function $a(t)$ is called the {\em cosmic scale factor}. Its role can be read off directly from the metric (\ref{FLRW-Metrik}): Assume that, at a fixed moment in time $t$, we determine the distance between two galaxies $A$ and $B$. Without loss of generality we can assume the y and z coordinates of these two galaxies to be equal (since we can always rotate our coordinate system so that two given galaxies $A$ and $B$ are separated in the x direction only). We obtain the distance between $A$ and $B$ by integrating the (infinitesimally) small distance element $\Dd s$ along the straight line joining $A$ and $B$, namely
\bea
\nonumber
d_{AB}(t) &=&\int\limits_A^B\Dd s = a(t)\,\int\limits_{x_A}^{x_B}|\Dd x| \\[0.5em]
&=& a(t)\cdot |x_B-x_A|.
\eea
Evidently, all distances between arbitrary pairs of galaxies in the Hubble flow change over time are proportional to the cosmic scale factor. For any two galaxies $A$ and $B$, and any two cosmic times $t_1$ and $t_2$, we have
\be
d_{AB}(t_1) = \frac{a(t_1)}{a(t_2)}\;d_{AB}(t_2).
\ee
This is the formula encoding what is meant by cosmic expansion: distances between distant galaxies changing proportionally to the same cosmic scale factor.

Next, we turn to gravitational waves: minute, propagating disturbances of spacetime geometry, travelling at the speed of light. Gravitational waves one can hope to detect are typically produced by the fast, accelerated movement of compact objects such as black holes or neutron stars.

The simplest metric for a gravitational wave in empty space can be written as
\begin{align}
\Dd s^2 = -c^2\,\Dd t^2 +  [1+h(t-z/c)] \Dd x^2 \notag\\
+[1-h(t-z/c)] \Dd y^2 +\Dd z^2.
\label{ppWelle-Metrik}
\end{align}
In the simplest case, the {\em gravitational wave strain} is a sine function
\be
h(t) = A\cdot\sin(\omega t),
\ee
possibly with an extra phase shift that is not included here, 
where the amplitude $A$ is extremely small for realistic gravitational wave signals, $A\sim 10^{-21}$. Similar to the cosmological case, the coordinates have been chosen to ensure each set of fixed spatial coordinates $(x,y,z)$ can be associated with a particle in free fall, which will keep its position in the coordinate systems even when a gravitational wave passes. (In other words, trajectories with $x=const., \; y=const.\; z=const.$ are geodesics, that is: possible trajectories for unconstrained particles in free fall.) In the absence of a gravitational wave, $h=0$, the coordinates reduce to the standard coordinates of special relativity.

The gravitational wave is propagating at the speed of light $c$; in the metric (\ref{ppWelle-Metrik}), propagation is in the positive z direction. The metric is a special case of a plane-fronted wave with parallel propagation, ``pp wave'' for short, derived from a linear approximation that treats gravitational waves as small deviations from an otherwise flat spacetime. The specific form given in  (\ref{ppWelle-Metrik}) corresponds to the transversal traceless gauge, ``TT gauge'' for short, and can be found in many university-level text books on general relativity.

\begin{figure*}[t]
\begin{center}
\includegraphics[width=1.7\columnwidth]{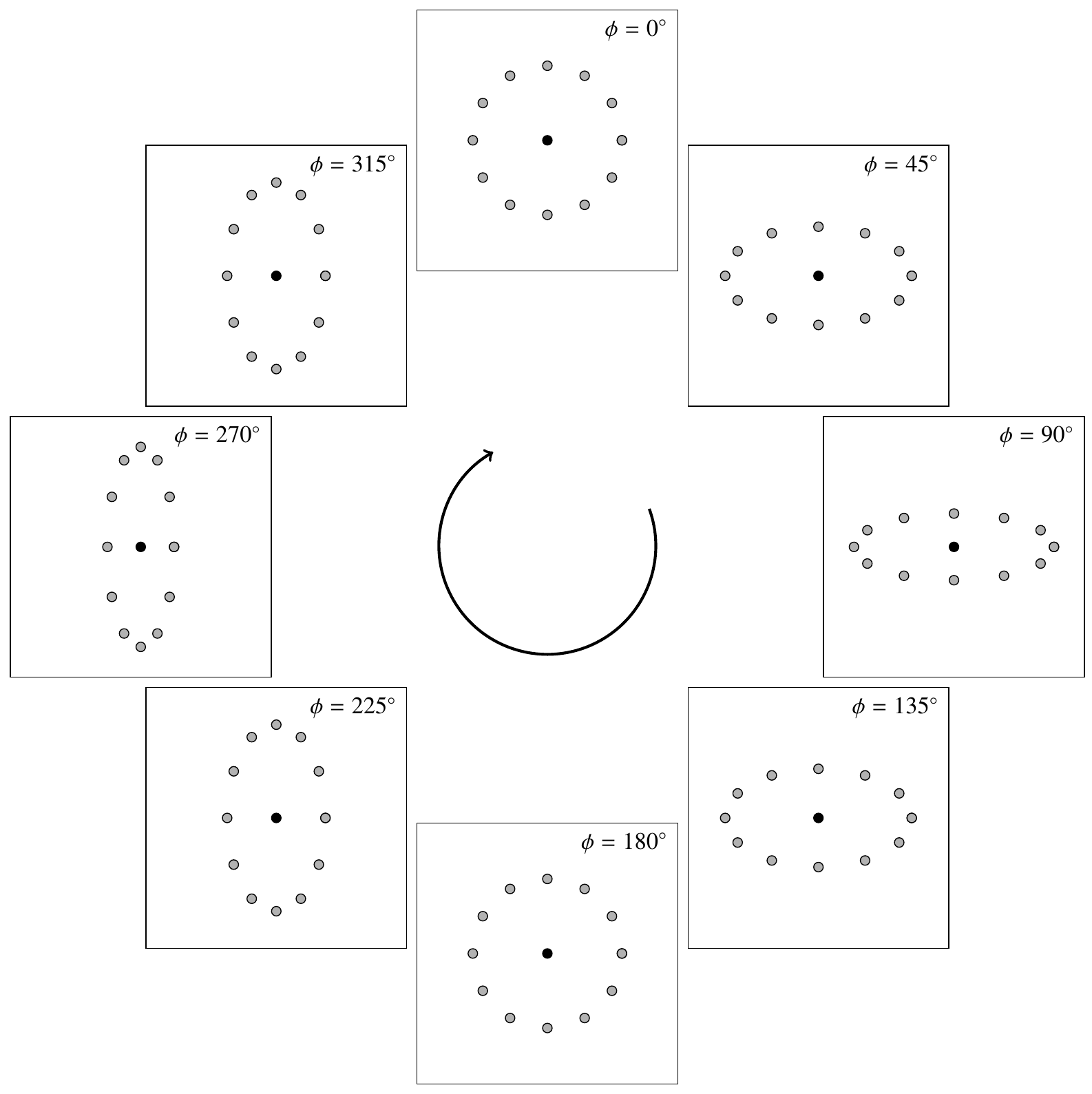}
\caption{Gravitational wave acting on a circle of freely floating particles. Plotted are the distances relative to a freely floating particle in the center of each image for various phases of the gravitational wave. The wave is propagating at right angles to the image plane, e.g. directly towards the viewer. The amplitude has been exaggerated by many orders of magnitude to render the effects of the wave visible. }
\label{GWCircle}
\end{center}
\end{figure*}

The action of the gravitational wave (\ref{ppWelle-Metrik}) can be seen most readily by examining a group of particles floating freely in the xy plane. Figure \ref{GWCircle} shows such a group of free particles, arranged to form a circle, at different phase values for the gravitational wave. The page corresponds to the xy plane, with the gravitational wave propagating from the back towards the observer. The pattern of stretching and shrinking, with maximal stretching in the horizontal direction while there is maximal shrinking in the vertical direction, and vice versa, is characteristic for gravitational waves, a direct consequence of what is meant when the waves are called quadrupole distortions. The direction of motion is perpendicular to the distortions, in other words: gravitational waves are transversal.

\subsubsection*{Similarities and differences}

Some similarities between the two metrics  (\ref{FLRW-Metrik}) and (\ref{ppWelle-Metrik}) are immediately obvious. In both cases, the deviations from flat spacetime (Minkowski metric) is in the spatial part only. In both cases, there are time-dependent metric coefficients in front of the terms $\Dd x^2$ and $\Dd y^2$ that do not depend on $x$ and $y$. 

Since the only term involving $\Dd t$ is, in both cases, $-c^2\Dd t^2$, the time coordinate is the time as shown by blocks that are at rest at constant coordinate locations. Also in both cases, constant coordinate locations are a form of free fall, in other words: In the absence of external non-gravitational forces, particles at rest in the spatial coordinates, floating in space, will remain floating at those same coordinate values of their own accord. 

On the other hand, there is a fundamental difference in that in cosmic expansion, all directions of space are on an equal footing, whereas a gravitational wave only affects the two spatial directions orthogonal to its direction of propagation. \columnbreak

In the xy plane, at constant $z$ and with $\Dd z=0$, we can write both metrics in the form 
\be
\Dd s^2 = -c^2\,\Dd t^2 + a_x(t)^2 \Dd x^2 +a_y(t)^2 \Dd y^2.
\label{AllgemeineMetrik}
\ee
From this generalized metric, we can derive some common effects that are present both in both spacetimes: in an expanding universe and as a gravitational wave passes by.  

\subsubsection{Frequency shift for light}
\label{FrequenzverschiebungLicht}

Consider light propagating in the xy plane with the general metric (\ref{AllgemeineMetrik}). More specifically, let us confine our attention to light propagating in one of the coordinate directions only, say: the $x$ direction. Using symmetry considerations, obvious for the cosmological case and somewhat more subtle in the case of the gravitational wave, it can be shown that this is no loss of generality.

The part of our metric dealing with the x and t directions only is
\be
\Dd s^2 = -c^2\,\Dd t^2 + a_x(t)^2 \Dd x^2.
\label{BeschraenkungMetrik}
\ee
Let $x(t)$ describe the propagation of light along the x direction. Per definition, light propagation means $\Dd s^2=0$, and we obtain
\be
c\frac{\Dd t}{a_x(t)} = \pm\Dd x.
\ee
The two different signs correspond to the two possible propagation directions of light, in the positive or negative x direction. 

Integrating up, we find that for light emitted at the location $x_e$ at the time $t_e$ and received at the location $x_r<x_e$ at the time $t_r$, we have
\be
c\int\limits_{t_e}^{t_r}\,\frac{\Dd t}{a_x(t)} = x_e-x_r.
\label{LichtAufintegriert}
\ee
Independent of the specific form of $a_x(t)$, we can derive the following: Consider {\em two} pulses of light emitted at the same location $x_e$ at times $t_e$ and $t_e+\Delta t_e$, which arrive at the location  $x_r<x_e$ at times $t_r$ and $t_r+\Delta t_r$. According to (\ref{LichtAufintegriert}), we have
\begin{eqnarray}
\nonumber
0&=&\int\limits_{t_e+\Delta t_e}^{t_r+\Delta t_r}\,\frac{\Dd t}{a_x(t)} -\int\limits_{t_e}^{t_r}\,\frac{\Dd t}{a_x(t)}\\[0.5em] 
\nonumber &=& \left[\int\limits^{t_r+\Delta t_r}_{t_r} + \int\limits^{t_r}_{t_e}-\int\limits^{t_e+\Delta t_e}_{t_e}
- \int\limits^{t_r}_{t_e}
\right]\frac{\Dd t}{a_x(t)}\\[0.5em] 
\nonumber &=& \left[\int\limits^{t_r+\Delta t_r}_{t_r} -\int\limits^{t_e+\Delta t_e}_{t_e}\right]\frac{\Dd t}{a_x(t)} \\[0.5em]
&\approx & \frac{\Delta t_e}{a_x(t_e)} - \frac{\Delta t_r}{a_x(t_r)}.
\label{RotverschiebungAbleitung}
\end{eqnarray}
The argument remains valid when we consider not light pulses, but consecutive wave crests of light waves. In this case, the $\Delta_t$ correspond to the oscillation period of the light, and are thus proportional to the light wave's wavelength $\lambda$. Thus, (\ref{RotverschiebungAbleitung}) entails
\be
\frac{\lambda_r}{\lambda_e} = \frac{a_x(t_r)}{a_x(t_e)}.
\ee
In an expanding universe, the scale factor has grown between the time $t_a$ and the later time $t_r$, so $a_x(t_r)>a_x(t_e)$. Thus, for distant galaxies whose light reaches us at the present time, the effect derived here corresponds to a systematic redshift known as the {\em cosmic redshift}.

On the other hand, light propagating at right angles to the direction of a gravitational wave is subject to a series of periodic red- and blue shifts that arise from the metric in exactly the same manner as the cosmic redshift. 

\subsubsection{Bound systems}

Next, consider a bound system in a spacetime described by the metric (\ref{AllgemeineMetrik}). This is more complicated than simply tracing the movement of particles in free fall (which amounts to finding the spacetime's straightest possible lines, or geodesics), as we need to take into account both gravitational acceleration and the non-gravitational forces responsible for keeping the system bound (or not). 

The most straightforward way of obtaining at least an approximate description of what happens to a bound system is based on Einstein's principle of equivalence, one of the fundamental principles of general relativity. The equivalence principle is the spacetime analogue of our zooming-in on an infinitesimal region of our curved surface discussed on page \pageref{CurvedSurface} and the following. In that case, a sufficiently high zoom factor made the region under scrutiny look indistinguishable from a subset of a Euclidean space. In fact, by a change of coordinates, we could have replaced the non-orthogonal coordinate system in figure \ref{ParallelogramGeo} by a proper orthogonal Cartesian system --- at least locally. The equivalence principle applies the same zoom-in principle to a general spacetime: At least locally, in the infinitesimal neighbourhood of any event, spacetime is indistinguishable from the flat spacetime of special relativity. By a suitable coordinate transformation, we can make that infinitesimal region look like flat spacetime in the usual coordinates, distances and light propagation described by the Minkowski metric (\ref{SRTMetrik}). There is a simple physical interpretation for the origin of such a locally Minkowskian system: the origin is the location of an observer in free fall. 

Let us choose just such a system; in fact, we can keep the origin to be the same as in our earlier version of the metric (\ref{AllgemeineMetrik}), given that the origin and any other fixed coordinate location correspond to the trajectories of particles in free fall. In this system, we will consider the non-relativistic limit (applicable to particles whose velocities are slow compared to $c$) and examine the consequences for a bound system described with the help of Newtonian physics.\footnote{This simplified pseudo-Newtonian derivation plus a more rigorous treatment can be found in \cite{Giulini2014,Carrera2008}}

For simplicity, we will again focus on one direction of space only, choosing once more the x direction, and the restricted metric (\ref{BeschraenkungMetrik}). One simple change is sufficient to make this metric look like that of special relativity at least locally: we introduce the new spatial coordinate $\bar{x} = a_x(t)\cdot x$, whose coordinate differences amount to spatial distances along the x axis, just as in the usual classical Cartesian  coordinate system. Direct calculation shows that, for a free particle with $x=const.$, we have 
\be
\ddot{\bar{x}} = \frac{\ddot{a}_x(t)}{a_x(t)}\cdot\bar{x}.
\ee
Through the lens of classical physics, this is an inertial acceleration affecting all particles equally. In systems with this kind of inertial acceleration, Newton's second law of mechanics takes on the modified form 
\be
m\left[\ddot{\bar{x}} - \frac{\ddot{a}_x(t)}{a_x(t)}\cdot\bar{x}\right] = F_x.
\label{MetrikKraftRadial}
\ee
In the case of a gravitational wave with the $a_x(t)$ defined in  (\ref{ppWelle-Metrik}), and neglecting higher-order terms in $A$, this equation can be written as
\be
m\left[\ddot{\bar{x}} + \frac{A\omega^2}{2}\cos(\omega[t-z/c]) \cdot\bar{x}\right] = F_x.
\label{GWErzwungen}
\ee
The simplest case is for the non-gravitational force $F_x$ to follow Hooke's law. The result is the simplest model for a so-called {\em resonant gravitational wave detector}: an oscillator with a forced sinusoidal oscillation. The detectors built from the 1960s onwards typically were metal cylinders with a length on the scale of a few meters, a meter in diameter, and with high quality factors to ensure that an oscillation excited by a passing gravitational wave following (\ref{GWErzwungen}) would have as large an amplitude as possible. These detectors were used in the first (and unsuccessful) attempts to directly detect gravitational waves.

The same formula can be applied to study the effects of cosmic expansion on a bound system. Instead of the more general form of the cosmic scale factor $a(t)$, we consider a Taylor expansion,  
\begin{align}
a(t) =\,&a_0 + H_0(t-t_0)\notag \\ 
+ &\frac12\, \alpha\, (t-t_0)^2 + O(3) 
\end{align}
where $H_0$ is known as the {\em Hubble constant}. For a bound system with Coulomb-like central force \be
F_c = -\frac{mC}{\bar{r}^2},
\ee
and with the realization that our formula (\ref{MetrikKraftRadial}) doesn't just apply to $\bar{x}$, but equally to a radial coordinate $\bar{r}$ measuring the distance from the origin, we have all the tools to model our system. As usual, our model includes a term depending on the angular momentum $L=\dot{\varphi}$ to make for an effective potential\footnote{This is described in more detail in \cite{Carrera2008}.
}
\be
\ddot{\bar{r}} = \alpha\cdot\bar{r} - \frac{C}{\bar{r}^2} + \frac{L^2}{R^3}.
\ee
Such a system will remain bound as long as 
\be
\bar{r} < \bar{r}_c = \left(\frac{C}{\alpha}\right)^{1/3}
\ee
with a critical radius $\bar{r}_c$. Using modern values for the cosmological parameters,\footnote{Based on measurements of the ESA satellite Planck and gravitational lensing data as documented in  \cite{Planck2013}.}  namely the Hubble constant $H_0 = 68\; \mbox{km}/(\mbox{s}\cdot \mbox{Mpc}) =
2.2\cdot 10^{-18}\;\mbox{s}{}^{-1}$ and the deceleration parameter $q_0 = -0.54$ (negative because cosmic expansion is, in fact, accelerating), we obtain 
\be 
\alpha = 2.66\cdot 10^{-36} \,\mbox{s}^{-2}.
\ee  
For a Coulomb system consisting of a proton and an electron, 
\be
C = e^2/(4\pi\varepsilon_0 m_e) = 253.27\,\mbox{m}^3/\mbox{s}^2,
\ee
 which corresponds to a critical radius $\bar{r}_c = 30{,}54\,\mbox{AU}$. Hence, a hydrogen atom is stable against the influence of (accelerated) cosmic expansion as long as the electron is less distant from its proton than Neptune is from the Sun! 

For the gravitational attraction acting on a planet that orbits the Sun, we have 
\be
C = GM_{\odot}\sim 10^{20}\;\mbox{m}^3/\mbox{s}^2,
\ee
corresponding to a critical radius of $\bar{r}_c =  390 $ light-years. A planetary orbit around the Sun is stable against the current (accelerated) cosmic expansion as long as it is smaller than that very large radius. 

Note that the only influence of cosmic expansion on bound systems is parametrized by the second-order term $\alpha$. Contrast this with popular accounts of the ``expansion of space,'' or worse,  ``new space being created as the universe expands,'' which suggest that the very fact that there is cosmic expansion should be enough to drive objects apart, say: to increase the distance between an electron in an atom and its proton. 

Instead, the situation exhibits the separation into kinematics and dynamics common in physical models: Dynamical laws do not fix velocities directly; instead, velocities -- e.g.\ the particular velocity field making up much of cosmic expansion -- are given as initial conditions. Dynamical effects make themselves felt via second-order, acceleration terms. Objects are not ``carried along'' by the kinematical part of cosmic expansion, parametrized by the Hubble constant; they experience the same acceleration acting on objects in the Hubble flow, and their motion changes accordingly.

\subsubsection*{Interferometric gravitational wave detectors}

With what we have learned in the preceding sections, we can also understand the basics of {\em interferometric gravitational wave detectors} such as those used for the first direct detection of gravitational waves in mid-September 2015, one-and-a-half weeks after our gravitational wave summer school in Jena.

\begin{figure}[H]
\begin{center}
\includegraphics[width=1.0\columnwidth]{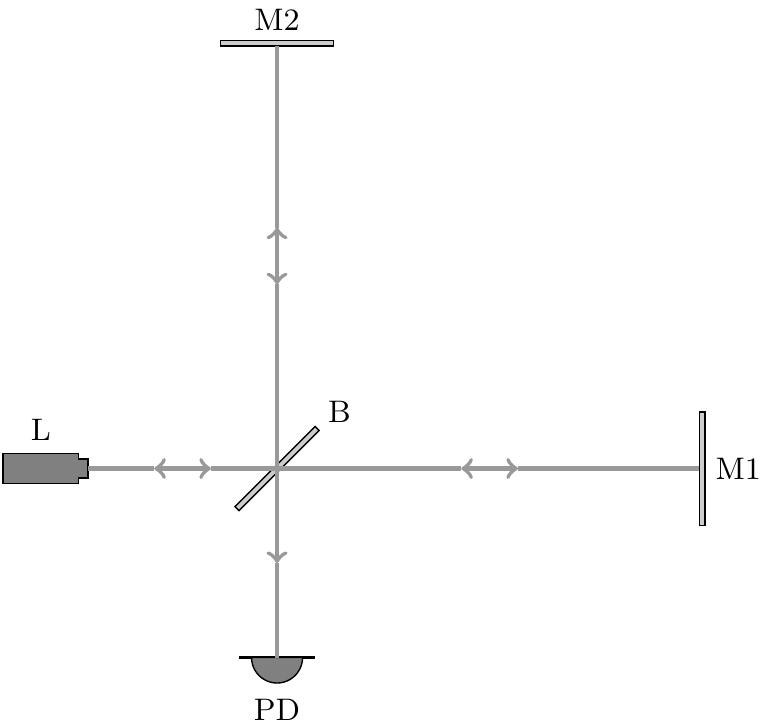}
\caption{Simple Michelson interferometer. Interferometric gravitational wave detectors are more complicated versions of this.}
\label{IntDetektor}
\end{center}
\end{figure}

The basic structure of such a detector is that of a Michelson interferometer, the simplest version of which is shown in figure \ref{IntDetektor}: Light from a laser source L propagates to the beam splitter B. Half of the light continues on to the mirror M1, while the other is reflected towards the mirror M2. After reflection at the respective mirrors, the light returns to the beam splitter B, where half of each portion moves on towards the photodetector PD. 

The mirrors M1 and M2 and the beam splitter B are suspended as multiple pendulums to isolate them as far as possible from external disturbances. When it comes to their reactions to a passing gravitational wave, and their motion back and forth along the directions of M1--B and M2--B, respectively, this makes them act approximately like particles in free fall. Interferometric detectors are commonly adjusted so that in the absence of a gravitational wave, hardly any light escapes towards the photodetector PD; in other words: there is almost complete destructive interference of the light moving from M1 and M2 in the direction of PD.

A gravitational wave passing through such a detector -- in the simplest case: orthogonal to the image plane of figure \ref{IntDetektor} -- makes a twofold difference. For one, the distances between the beam splitter and the end mirrors will change in the same way as the distances between the freely falling particles in figure \ref{GWCircle}; in the simplest case, one arm will be stretched while at the same time the other arm will be shrunk. Destructive interference cannot be maintained under such conditions, and light will leak out towards the photodetector. After all, the time it takes for wave crests and troughs via M1 and M2 to the detector PD will change as the relative armlengths change.

An additional effect are the frequency shifts for light, discussed at page  \pageref{FrequenzverschiebungLicht} and the following. As arms are stretched and shrunk, light within the detector is red- and blueshifted accordingly, proportional to the relevant scale factor. To light waves that have different wave lengths can never have complete destructive interference; this effect contributes to the light leaking out at the photodetector, as well. When the time light remains inside the detector is short compared to the oscillation period of the gravitational wave, as in current ground-based detectors, the consequences of this effect are much smaller than the change in armlengths. For much longer light-travel times, such as in the different varieties of the proposed space-based detector LISA, gravitational-wave-induced frequency shifts become more important.\footnote{A helpful discussion of this can be found in \cite{Saulson1997}.}

\subsubsection*{Conclusions}

The general-relativistic descriptions of cosmic expansion and of gravitational waves have similarities that are helpful in understanding both phenomena. In their simplest incarnations, both feature a scale factor or multiple scale factors in front of an otherwise flat spatial metric. In cosmology, this scale factor governs the changing distances of particles in the Hubble flow, in the case of gravitational waves the changing distances between various components of an interferometric detector. 

When non-gravitational forces are present, it is easiest to change to a pseudo-Newtonian picture. Here, the second derivatives of the scale factors cause inertial accelerations that can be contrasted with the acceleration a particle experiences through a Coulomb-like force. In the cosmological case, this helps to distinguish kinematic and dynamic components, showing that particles are not ``whisked along'' with the Hubble flow, but given arbitrary initial conditions, react only to the second-order, dynamical influence of scale-factor expansion. In the case of gravitational waves, the corresponding description allows for an understanding of resonant wave detectors. 

A spacetime metric directly governs the propagation of light, and the scale-factor metrics studied here allow for simple calculations of the wavelength shifts experienced by light. In the cosmological  case, this yields the famous cosmological redshift. For gravitational wave, it yields red- and blueshifts that become important for future large-scale, space-based detectors.

\subsubsection*{Acknowledgements}

I am grateful to B. Schutz for discussions during the summer school that proved helpful for this text and corresponding lecture, and to T. M\"uller for comments on a draft version of this text. 
\end{multicols}

\subsubsection*{References} 
\begingroup
\renewcommand{\chapter}[2]{}%

\endgroup
\clearpage

\end{document}